\documentclass[11pt]{article} 
\pdfoutput=1
\begin{document} 
\title{Vortex Merger of Long Filaments}
\author{Akshay Khandekar and Jamey Jacob \\ \\\vspace{6pt} School of Mechanical and Aerospace Engineering, \\ Oklahoma State University, Stillwater, OK, USA 74078}
\maketitle
\begin{abstract} Vortex merger of long filaments. \end{abstract}
\section{Introduction}
This fluid dynamics video demonstrates the merger of long vortex filaments is shown experimentally. Two counter-rotating vortices are generated using in a tank with very high aspect ratio. PIV demonstrates the merger of the vortices within a single orbit. 
\end{document}